\documentclass[aps,preprint,amssymb,preprintnumbers,amsmath,nofootinbib]{revtex4-1}

\usepackage{graphicx}

\newcommand{\be}{\begin{equation}}
\newcommand{\ee}{\end{equation}}
\newcommand{\ba}{\begin{eqnarray}}
\newcommand{\ea}{\end{eqnarray}}

\newcommand{\beq}{\begin{equation}}
\newcommand{\eeq}{\end{equation}}

\newcommand{\Eq}[1]{Eq.~\eqref{#1}}

\newcommand{\hq}{{\hat q}}
\newcommand{\Q}{{q_0,{q}}}

\usepackage{slashed}
\usepackage{hyperref}

\date{\today}

\begin{document}

\title{Damping rate of a fermion in ultradegenerate chiral matter} 

\author{Stefano Carignano}
\affiliation{Departament de F\'\i sica Qu\`antica i Astrof\'\i sica 
                   and Institut de Ci\`encies del Cosmos,
        Universitat de Barcelona,
        Mart\'\i $\,$ i Franqu\`es 1, 08028 Barcelona, Catalonia, Spain.}

\author{Cristina Manuel}
\affiliation{Instituto de Ciencias del Espacio (ICE, CSIC) \\
C. Can Magrans s.n., 08193 Cerdanyola del Vall\`es, Catalonia, Spain
and \\
 Institut d'Estudis Espacials de Catalunya (IEEC) \\
 C. Gran Capit\`a 2-4, Ed. Nexus, 08034 Barcelona, Spain
}

\begin{abstract}
We compute the  damping rate  of a fermion propagating in a chiral plasma when there is  an imbalance between the densities of left- and right-handed fermions, after generalizing
the hard thermal loop resummation techniques for these systems.
In the ultradegenerate limit, for very high energies the damping rate of this external fermion approaches a constant value. 
Closer to the two Fermi surfaces, however, we find that the rate depends on both the energy and the
chirality of the fermion, being higher for the predominant chirality.
 This comes out as a result
of its scattering 
with the particles of the plasma, mediated by the exchange of
Landau damped photons. 
 In particular, we  find that the chiral imbalance is responsible for a different propagation of  the left and right circular polarised transverse modes of the photon, 
and that a chiral fermion interacts differently with these two transverse modes.
 We argue that spontaneous radiation
of energetic fermions is kinematically forbidden, and discuss the time regime where our computation is valid.

\end{abstract}

\maketitle

\section{Introduction}

The study of chiral matter characterized by different
 densities of right-handed and left-handed massless fermions 
has received increasing attention in last years. This is due to its  wide range of applications,
 from the description of the quark-gluon plasma produced in relativistic heavy ion collisions to the
 so-called Weyl and Dirac semi-metals in condensed matter physics, as well as its potential relevance for cosmological and astrophysical scenarios \cite{Kharzeev:2013ffa,Kharzeev:2015znc,Huang:2015oca,Landsteiner:2016led,Gorbar:2017lnp,Joyce:1997uy,Tashiro:2012}.  Systems of this type exhibit a number of unique features of extreme interest.
 Novel transport phenomena associated with the quantum chiral anomaly are manifested at 
 a macroscopic level \cite{Vilenkin:1980fu,Vilenkin:1979ui,Fukushima:2008xe,Son:2009tf}. Further, the breaking of parity $P$ and $CP$, where $C$ stands for charge
 conjugation, has also a clear effect on the quasiparticles and collective modes propagating in the plasma.

In this manuscript we focus our attention on  the damping rate of a  fermion propagating in  chiral matter,  whose inverse can typically be interpreted as the quasiparticle lifetime.
As it is already known for the similar case of a plasma without chiral imbalance,  at finite temperature the damping rate turns out to be
infrared  divergent even after taking into account screening corrections of the electromagnetic interactions, and only non-perturbative approaches such as that of Ref.~\cite{Blaizot:1996hd} lead to  a non-exponential fermion
decay law. In this work we will instead focus on the case of a vanishing temperature but finite density. In this case the 
result for the damping rate is instead finite, and for the case of a regular ultradegenerate plasma 
 the relativistic Fermi sea was found to be stable \cite{LeBellac:1996kr,Manuel:2000mk,Vanderheyden:1996bw}. 
The magnetic  unscreened interaction nevertheless leads to a non-Fermi liquid behavior of the relativistic plasma \cite{Holstein,Manuel:2000mk}.  

Let us now turn to the specific case of a plasma with a chiral imbalance. As one might expect, in this case the damping rate for a fermion will depend on its chirality. First of all, the
 fermion scatters with the particles in the plasma, subject to the Pauli blocking
effects, which are different depending on its chirality. But it is also important to note that the chiral imbalance
affects the photon mediated interactions, by making that left and right handed  circular polarized transverse photon modes propagate
differently than in a medium without chiral imbalance. 
In the following, we will show  this explicitly by extending the  resummation methods  developed for the study of QCD and QED relativistic plasmas \cite{Braaten:1989mz,Bellac:2011kqa} to   systems with chiral imbalance. 
Our explicit  computation also shows that a chiral fermion interacts differently with the two circular polarized photons, 
 a novel effect that, to the best of our knowledge, has never been discussed in the literature.

Furthermore,  resummed propagators in chiral plasmas contain purely imaginary poles associated with the transverse photon modes, 
leading to an instability which results in an exponential growth of the associated gauge fields \cite{Akamatsu:2013pjd}.  
This is similar to the appearance of instabilities in spatially anisotropic system, a topic which has been discussed at length the context of heavy-ion collisions \cite{Mrowczynski:2016etf}.
Accordingly, the time scale associated with the exponential growth of the fields will delimit the region in which perturbative calculations like the ones discussed in this work are applicable.

 This paper is organized as follows. In Sec. \ref{sec:HTL} we show how to extend  the Hard Thermal Loop (HTL) resummation scheme in plasmas with chiral imbalance. The
most relevant effect comes in the splitting of the transverse modes into left and right handed circular polarized modes. We present the dispersion modes associated with two stable
collective modes, together with their corresponding spectral functions. We also discuss the presence of unstable gauge modes.
 In Sec. \ref{sec:damping} we apply this extended resummation program to he computation of the damping rate of a fermion in a chiral imbalanced plasma at vanishing temperature. We finally discuss our conclusions in Sec. \ref{conclusions}. We use natural units ($\hbar = c = k_B=1$) throughout, and we denote by $e = \sqrt{4\pi\alpha}$ the QED coupling constant.

\section{HTL and collective modes for a chiral plasma} 
\label{sec:HTL}

 Let us consider an energetic fermion crossing a chiral QED plasma, 
characterized 
 by having different chemical potentials associated with left-handed and
right-handed  populations of fermions, $\mu_L \neq \mu_R$. 
We introduce a chiral chemical
potential, defined as $\mu_5 = \frac 12 (\mu_R - \mu_L)$, to quantify this imbalance, and assume it without loss of generality to be positive. 
We also consider the presence of a vectorial chemical potential $\mu_V > \mu_5$, so that $\mu_{R/L} = \mu_V \pm \mu_5$ are both positive.

Aside from the direct effect on the fermion propagator due to the presence of two Fermi surfaces (right and left), 
chiral imbalance also affects the photon propagator entering in the one-loop fermion self-energy due to the breaking of $P$ 
and $CP$ \cite{Nieves:1988qz}. 
It is known that the leading contribution to the damping rate when the fermion energy $E$ is large
comes from values of the momenta which are soft, or of the order of Debye mass  \cite{Blaizot:1996hd,LeBellac:1996kr,Manuel:2000mk}, and thus resummed 
hard thermal/dense loop 
propagators have to be used \cite{Bellac:2011kqa}. In a chiral plasma we see that the two transverse photon propagators split.
In  Coulomb gauge and working in the medium rest frame 
(we ignore gauge dependent pieces here, as they do not contribute to the imaginary part of the fermion self-energy)
the photon propagator can be written as  \cite{Nieves:1988qz}
\be
\Delta_{\mu \nu} (q_0,{\bf q}) = \delta_{\mu 0} \delta_{\nu 0} \Delta_L(\Q) + \sum_{h=\pm} {\cal P}^{T,h}_{\mu \nu} \Delta_T^h(\Q) \,,
\ee
where $q=\vert {\bf q} \vert$ and  $h=\pm$ labels the two circular polarised transverse photon states, left and right,  and we introduced  
\be
{\cal P}_{\mu \nu}^{T,h} = \frac 12 \Big(\delta^{ij} - \hq^i \hq^j - i h \epsilon^{ijk} \hq^k \Big) \delta_{\mu i} \delta_{\nu j} \ .
\ee
The resummed longitudinal and transverse propagators  read, with the usual prescription  $q_0 \rightarrow q_0 \pm i \eta$ for retarded and advanced quantities, respectively,
\be
\Delta_L (q_0, q)= \frac{1}{q^2 + \Pi_L} \ , \qquad  \Delta_T^h(q_0,q) = \frac{1}{q_0^2 -q^2 - \Pi_T - h \Pi_P}\,,
\ee
where 
\begin{eqnarray}
\label{pipiL}
\Pi_L (q_0,q) & = & m^2 _{D}  \left(1- \frac{q_0}{2 q} 
 \,{\rm ln\,}{\frac{q_0+ q}{q_0- q}}  \right)
  \ , \\
 \Pi_T (q_0,q) & = & m^2 _{D} \, \frac{q_0^2}{2  q^2} \left[ 1 + \frac12 \left( \frac{q}{q_0} -
\frac{q_0}{ q} \right) \,  {\rm ln\,} {\frac{q_0+
 q}{q_0- q}} 
\, \right] \ ,
 \label{pipiT}
\end{eqnarray}
are the longitudinal/transverse part of the hard thermal/dense loop photon polarization tensor \cite{LeBellac:1996kr,Manuel:2000mk},
and
 \be
 m^2_D = e^2  \left( \frac{T^2}{3} + \frac{\mu_R^2 + \mu_L^2}{2 \pi^2} \right)
\ee
is the Debye mass, 
while 
\be
\Pi_P (\Q)=  - \frac{e^2 \mu_5}{2 \pi^2}  \frac{q_0^2-q^2}{ q } \Big[ 1 - \frac{q_0}{2  q}   {\rm ln\,} {\frac{q_0+ q}{q_0-q}}
  \Big] \,
\ee  
can be viewed as the anomalous hard dense loop contribution \cite{Laine:2005bt,Akamatsu:2013pjd,Manuel:2013zaa}.

The spectral functions associated with the gauge field modes
are given by
\ba
 \rho_{L} (\Q) & =  &2 \,{\rm Im} \,\Delta_{L}(q_0 + i\eta, q) \ ,  \\
  \rho_{T}^h (\Q) & = &  2 \,{\rm Im} \,\Delta_{T}^h(q_0 + i\eta, q)  \ , \qquad h= \pm \ .
\label{spectral}
\ea
The spectral
functions  for the resummed photon propagators   for $\mu_5 =0$
can be found in \cite{Bellac:2011kqa} (see also \cite{Manuel:2000mk}), and we now generalize them for the
case when there is chiral imbalance.
As customary, they can be conveniently split into two parts,   $\rho(\Q) = \rho^>(\Q) + \rho^<(\Q)$,
 the former being only non-vanishing
above the light-cone, while
the latter (associated with Landau damping) is nonzero only for space-like momenta.
The longitudinal spectral function is the same as in a chiral symmetric plasma, while for the two transverse modes ($h= \pm$) one finds
\beq
\frac{\rho^{h,>}_{T} (\Q)}{2 \pi} = Z^h_{T} 
\left[\delta(q_0 - \omega^h_{T} (q)) -
\delta(q_0 + \omega^h_{T} (q)) \right] \ ,
\eeq
where  $\omega^h_{T}$  are the real poles of $\Delta_{T,h}$,  given by solving for $q_0$ the equation
\be
\label{eq:polesreal}
\left(q_0^2 - q^2\right) \left[ 1 - \frac{1}{2q^2}\left(m_D^2 + hCq \right)\left(1-\frac{q_0}{2q}  {\rm ln\,}  \frac{q_0 + q}{q_0 - q}  \right) \right] - \frac{m_D^2}{2} = 0
\ee 
and $Z^h_T$
  are the corresponding residues, given by
  \be 
Z^h_T (k) = \frac{ \omega_{T,h} ( \omega^2_{T,h} -k^2)}{( \omega^2_{T,h} -k^2)^2  -m^2_D \omega^2_{T,h}  - \frac{hCk}{2}  ( \omega^2_{T,h} -k^2) } \,,
\ee
where 
\be
C =  -\frac{e^2 \mu_5}{\pi^2} 
\ee
 measures the chiral imbalance.
 The remaining part of the spectral function for values below the light-cone is given by
 \beq
\label{eq:rhocut}
\frac{\rho^{h,<}_{T}}{2\pi}  =
 \frac{ M_h^2 \, \, \frac{x}{1-x^2} \, \Theta(1-x^2)}{
\left[ 2 q^2 +\frac{m_D^2}{1-x^2} - M_h^2 Q_1(x) \right]^2 + M_h^4 \frac{\pi^2 x^2}{4}} \ , \qquad x =q_0/ q , 
\eeq
with $M_h^2 =  m_D^2 + h C q$, where  
$
Q_1(x) = 1 - \frac{x}{2} {\rm ln\,} \left|{\frac{1+x}{1-x}}\right|  $.

 There are both collective stable and unstable gauge modes in the chiral plasma.  
  The real poles of $\Delta_{T,h}$
 describe the existence of two different transverse stable collective modes  identified by their  circular polarization $h$.
For small values of $q$, $q \ll m_D$, one finds the approximate behavior 
\be
\omega^2_{T,h} \approx  \omega^2_p  + \frac 65 k^2 + \frac{hC}{6} k \,,
\label{eq:smallq}
\ee
where $\omega_p=m_D/\sqrt{3}$ is the plasma frequency, 
while for $q \gg m_D$ one gets 
\be
\omega^2_{T,h} \approx k^2 + \frac 12 m^2_D  + h C\frac{m^2_D}{4 k} \left(1 - \frac 12 \ln{\frac{8 k^2_D}{m_D^2}} \right) \,,
\label{eq:largeq}
\ee
both of which reduce to the known expressions \cite{Bellac:2011kqa} when $C\to 0$.

\begin{center}
\begin{figure}
\includegraphics[width=.4\textwidth]{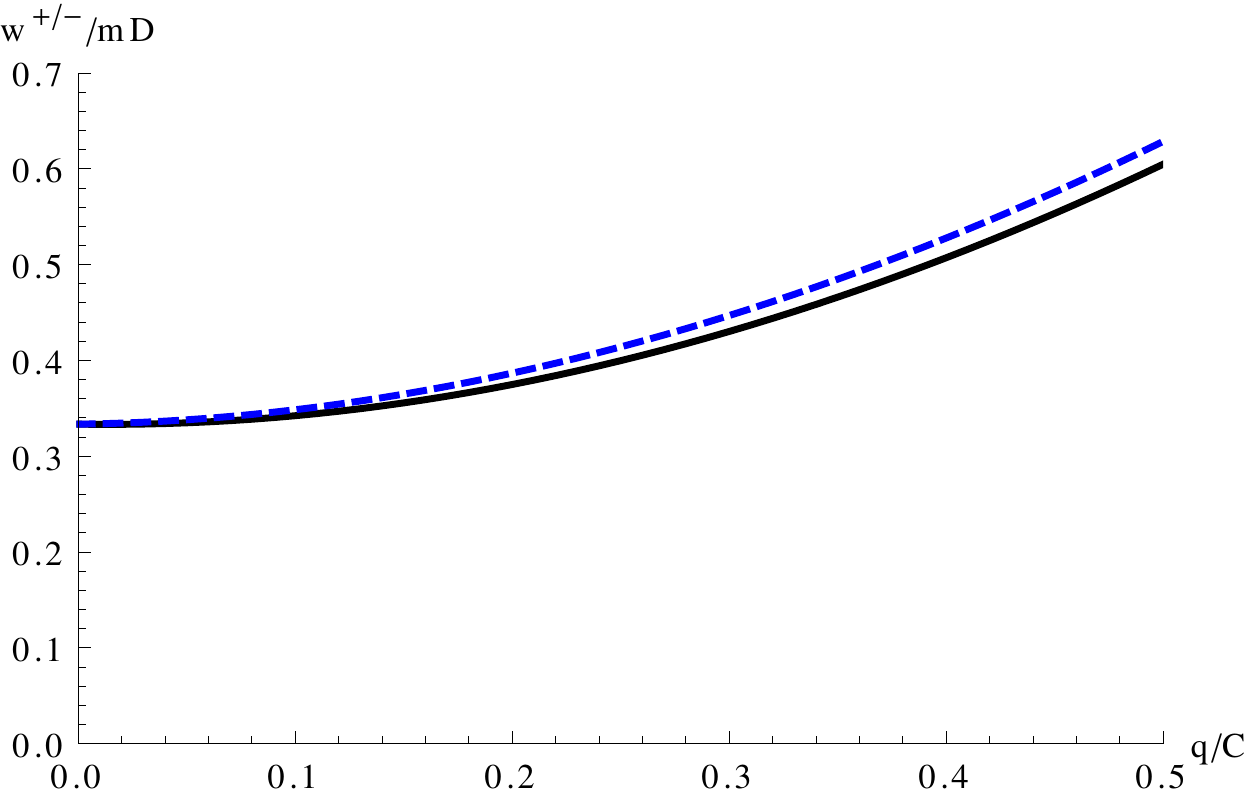}
\includegraphics[width=.4\textwidth]{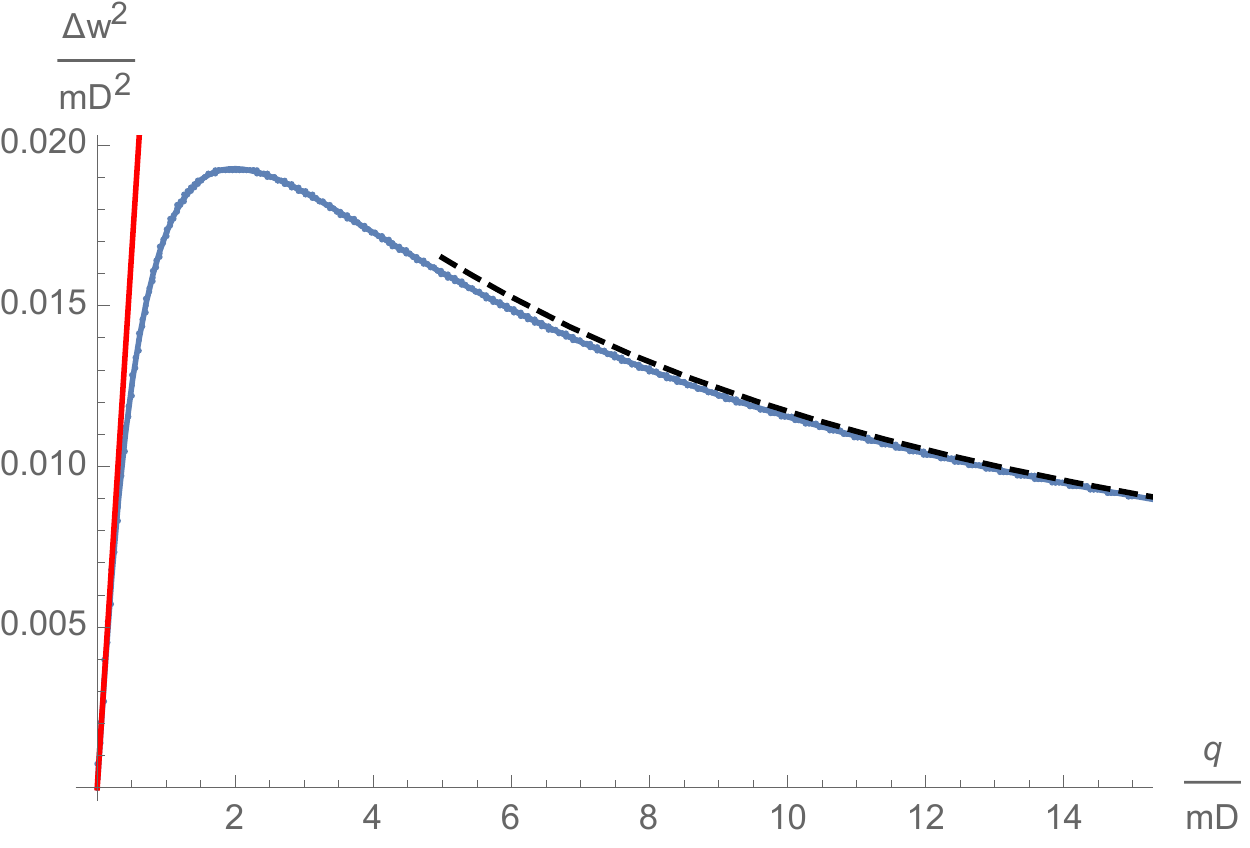}
\caption{   Left: full numerical result for the real poles $\omega_{T,\pm}$ associated with the different photon polarizations as function of $q$ in units of $C$, showing the increase of the splitting between the two modes. The black solid line denotes the result for the positive helicity, while the blue dotted refers to the negative helicity.
 Right: difference $\Delta\omega^2 = \omega^2_{T,+} - \omega^2_{T,-}$ as function of the momentum $q$ in units of $m_D$. The red line denotes the analytical result for the difference computed for small $q$ from \Eq{eq:smallq}, while the black dashed one denotes the analytical result from large $q$ from \Eq{eq:largeq}. All these results are plotted for a fixed value $C = 0.15 m_D$.
  \label{fig:poles} }
\end{figure}
\end{center}

 In Fig. \ref{fig:poles} we show the full numerical result for the real poles of $\Delta_{T}^{h}$ for the two different polarizations. The two modes are degenerate at $q=0$ and their splitting increases as $q$ grows,
up to a maximum value which is reached around $q \sim 2 m_D$, after which it starts decreasing again.   

  Aside from these real solutions, it is possible to find an additional family of poles of $\Delta_{T}^h$ which are purely imaginary.  Indeed, for 
$q_0 \ll  q$ one can expand the denominator of $\Delta_{T}^h$ as 
\beq
q_0^2 -q^2 = \pm \frac{e^2 \mu_5}{2\pi^2}  q - i \frac {\pi m^2_D}{2} \frac{q_0}{q} \,,
\eeq
 from which one gets 
\be
q^0 =- \frac{2 i}{\pi m_D^2} q^2 \left(q - \frac{hC}{2} \right) 
\ee
These purely imaginary poles give rise to the chiral plasma instability discussed in Ref.~\cite{Akamatsu:2013pjd}.

We have checked numerically that no other families of complex solutions are present.

Let us also mention that the  dispersion law (without the imaginary term) was employed in  Ref.~\cite{Tuchin:2018sqe}.
Our computations show that this is only valid for space-like momenta, so that it cannot lead to processes with spontaneous radiation. 
Let us stress that the low-energy theory used in Ref.~\cite{Tuchin:2018sqe} to describe chiral matter is only strictly valid in the static limit, see also \cite{Laine:2005bt,Manuel:2013zaa}.

\section{Fermion Damping rate} 
\label{sec:damping}

We now consider a specific application of the perturbation theory developed in the previous section: the damping rate of a fermion in a chiral plasma. This 
can be obtained by explicitly computing the Feynman diagram associated with the scattering between the fermion and the plasma particles, or equivalently from
  the imaginary part of the retarded fermion self-energy $\Sigma$ evaluated on-shell \cite{Weldon:1983jn}.
As anticipated, in a chiral medium this quantity depends on its chirality $\chi$, and it can be expressed as
\be
\gamma_{\chi}(E) = -\frac{1}{2E} {\rm Tr} \Big[ {\cal P}_{\chi} \slashed{p} \,  {\rm Im} \Sigma(p_0 + i \eta, {\bf p}) \Big] \Big \vert_{p^0=E} \,,\; \eta \rightarrow 0^+ \, 
\ee
where  ${\cal P}_{\chi} = \frac{1}{2} \left(1 + \chi \gamma_5\right)$ are the chiral projectors, with the assignment $\chi = \pm1$ for right/left-handed  fermions, respectively.
\begin{center}
\begin{figure}
\includegraphics[width=.4\textwidth]{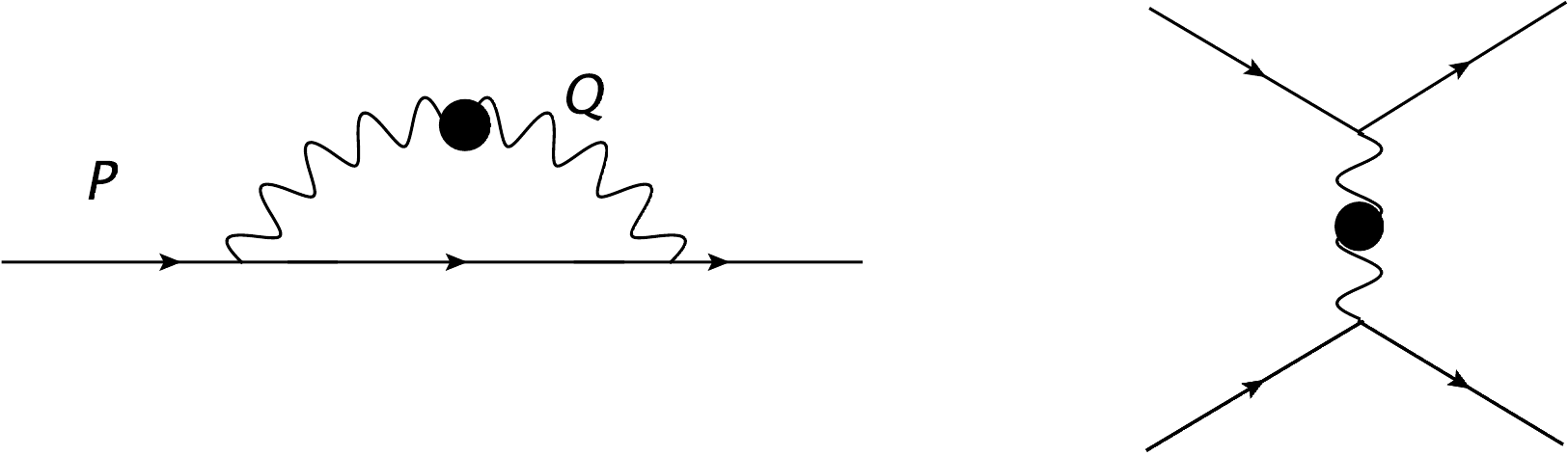}
\caption{ Left: Fermion self-energy entering the equation for the damping rate. Right: corresponding scattering process at $T=0$: the energetic fermion interacts with particles from the medium via the exchange of  Landau damped photons.  The dot denotes a resummed propagator. \label{fig:self} }
\end{figure}
\end{center}
 The computation of the fermion self-energy (Fig.\ref{fig:self}, left)  can be carried out,  {\it e.g.}, in the imaginary time formalism of thermal field theory and then analytically continued to Minkowski space.
It is the same sort of computation as the one carried out in Refs.
\cite{LeBellac:1996kr,Manuel:2000mk}, although as we will see the presence of a chiral imbalance leads to an asymmetry
 between the left and right-handed components of the fermion self-energy.

By following the same initial steps as in \cite{LeBellac:1996kr} while taking into account the new structure of the photon propagator in a chiral plasma,
the damping rate can be expressed in terms of the fermion and photon spectral functions as
\begin{align}
\label{basic}
\gamma_{\chi}(E) 
&= -\frac{e^2}{2E} \, {\rm Im}\, 
 \int{{\rm d}^4q\over (2\pi)^4} \int^{\infty}_{-\infty}
\frac{{\rm d} k_0}{2 \pi} \rho_f(k_0,k)  
 \frac{1 + f(q_0) - {\tilde f}(k_0 -\mu_\chi)}{p_0 -k_0 -q_0 + i \eta}  {\rm Tr} \left({\cal P}_{\chi} \slashed{p} \gamma^\mu  \slashed{k}  \gamma^\nu \right) 
\nonumber \\
&\times \left[\delta_{\mu 0}
 \delta_{\nu 0}
\rho_L(\Q) +  \sum_{h=\pm} {\cal P}^{T,h}_{\mu \nu} \rho^h_T(\Q)\right] \Bigg\vert_{p^0=E} \ ,
\end{align}
where $k^\mu = (k^0, {\bf k} = {\bf p} - {\bf q})$
where
 \be
 \rho_f(k_0,{\bf k}) = 2\pi {\rm sgn}(k_0) \delta(k_0^2 - |{\bf k}|^2)
 \ee
 is the fermion spectral function, and  $f$ and ${\tilde f}$ are Bose-Einstein and Fermi-Dirac
distribution functions ($\beta=1/T$), respectively
\begin{equation}
f(q_0)={1\over {\rm e}^{\beta q_0}-1} \ , \qquad
{\tilde f} (k_0 -\mu_\chi)={1\over {\rm e}^{\beta (k_0-\mu_\chi)}+1} \ .
\label{eq17}
\end{equation}

Eq.~(\ref{basic}) essentially gives account of scattering  (see Fig.1,right) and/or annihilation processes
of the external fermion with the particles of the medium, through the exchange of  photons.

In the following we will focus on the study of the ultradegenerate limit and set $T=0$, as in this case the imaginary part of the fermion self-energy
is finite, after taking account the resummation in the photon propagators. While at finite $T$ both fermion-fermion scattering and fermion-antifermion annihilation
processes contribute to the damping, at $T=0$ only the first process is possible.
Let us first concentrate on the value of the damping rate when the photon momentum
is soft (that is, of the order of the Debye mass) and introduce an intermediate cutoff $q^*$, such that $m_D \ll q^* \ll m_D/e$. 
 We also assume that $E \geq \mu_L, \mu_R$. Then, 
 expanding $ |{\bf k}| = E\sqrt{1  -  2q \cos\theta/E + q^2/E^2}$ up to order $1/E$
  one reaches

\begin{align}
\label{generalexp}
\gamma^{\rm soft}_{\chi} & \simeq  \frac{e^2}{2}  \int \frac{d^3q}{(2\pi)^{3}} \Big (\Theta(q_0) - \Theta(\mu_{\chi} -E+ q_0) \Big) \Theta(q^* -q) 
\nonumber \\
 \times & \Bigg\{ \rho_L (\Q) \Big(1 - \frac{q_0}{E} \Big)  +\frac 12 (1 -\cos^2{\theta}) \sum_{h=\pm} \Big [ \Big(1 - \frac{q_0}{E} - \chi h \frac{q}{E} \Big) \rho_{T}^h(\Q)  \Big] \Bigg\}  \Bigg |_{q_0 = q\cos\theta} \,,
\end{align}
where we have kept terms up to order $1/E$. 
For $\mu_5=0$, the leading term gives the same result obtained in Ref.~\cite{LeBellac:1996kr}
  as then $\mu_R = \mu_L$, and also $\rho^+_{T}  = \rho^-_{T} $. 
  In presence of chiral asymmetry, however, the leading term in the energy expansion
 still depends on the chirality of the fermion involved, both through the Pauli blocking effects encoded in $\mu_\chi$ and the $\mu_5$ dependence in the transverse photon spectral functions.
 At this point we may already notice upon inspecting Eq.~(\ref{generalexp}) that at the Fermi surface, $E = \mu_\chi$, the damping rate vanishes.  This is simply due to Pauli blocking, which is enforced by the occupation numbers which at vanishing temperature reduce to sharp step functions. 
 At zero temperature the 
 damping rate is thus found to be zero even in presence of a chiral imbalance.
 Of particular interest is the last term of Eq.~(\ref{generalexp}), which shows that a fermion of definite chirality interacts differently with the transverse photons depending on their polarization.
 While this effect is of order $1/E$, it provides a qualitative new feature which may play an important role in chiral plasmas, and deserves to be better explored.
  It is possible to derive the associated different $1/E$ vertices of these couplings in a systematic way using the 
  On-Shell Effective Field Theory (OSEFT) proposed in \cite{Manuel:2014dza}.
  
  We have checked that the expression given in Eq.~(\ref{generalexp}) corresponds to the decay rate that might be computed by considering the scattering process
  of Fig. 2 right , at leading order in the energy expansion, following a similar analysis as the one carried out in Ref.~\cite{Blaizot:1996az} but here at zero temperature.
  
  In the imaginary part of the fermion self-energy, it is only $\rho^<$ which contains effects  Landau damping part 
that contributes, as the kinematical constraints
force the photon momentum to be space-like $(q_0^2 \leq q^2)$.  
 After plugging in \Eq{generalexp} the expressions derived in the previous section for the spectral functions, we arrive at 
\begin{widetext}
\begin{align}
 \label{eq:damping0}
\gamma^{\rm soft}_{\chi}(E) & =  \frac{e^2}{8 \pi } \int_{D_{\chi}} dq_0 dq \Bigg \lbrace \frac{q_0 m_D^2}{\left[q^2 + m^2_D Q_1(x) \right]^2 + \frac{\pi^2m^4_D x^2}{4}} 
 \left( 1 - \frac{q_0}{E} \right)
 \\
\nonumber
 +  \sum_{h=\pm} &  \left[ \frac{q^0 (  m_D^2 + hCq) }{\frac{\pi^2 x^2}{4} (hC q + m_D^2)^2 + \left[ 2 q^2 -(h Cq+m_D^2) Q_1(x) + \frac{m^2_D}{1-x^2} \right]^2} 
 \Big( 1 - \frac{q_0}{E} - \chi h \frac{q}{E} \Big) \right] 
 \Bigg \rbrace + {\cal O}\Big(\frac{1}{E^2}\Big) \,,
 \end{align}
  \end{widetext}
where
 the domain of integration is
\be
D_{\chi} = \left \{ 0 \leq q_0 \leq E -\mu_{\chi} \, ; q_0 \leq q \leq q* \right \} \,.
\ee
The hard contribution to the damping rate can be estimated from Eq.~(\ref{generalexp})  using free or unresummed propagators (that is, setting $m_D =C=0$ in the spectral functions)
and taking into account that the momentum transfer should obey $q > q*$, and $q < q_{\rm max} = 2 E$, where $q_{\rm max}$ is the maximum momentum transfer
allowed by the kinematics. 

 Let us remark here that  by working at zero temperature we do not have the characteristic Bose enhancement at vanishing frequencies, and the presence of unstable modes does not lead to problems of non-integrability or ill-behaved terms in the above integrals.
 This is similar to what occurs in the computation of energy loss in systems with Weibel instabilities,
 where one witnesses a sort of effective ``dynamical shielding'' of the unstable modes \cite{Romatschke:2003vc}.

\begin{center}
\begin{figure}
\includegraphics[width=.4\textwidth]{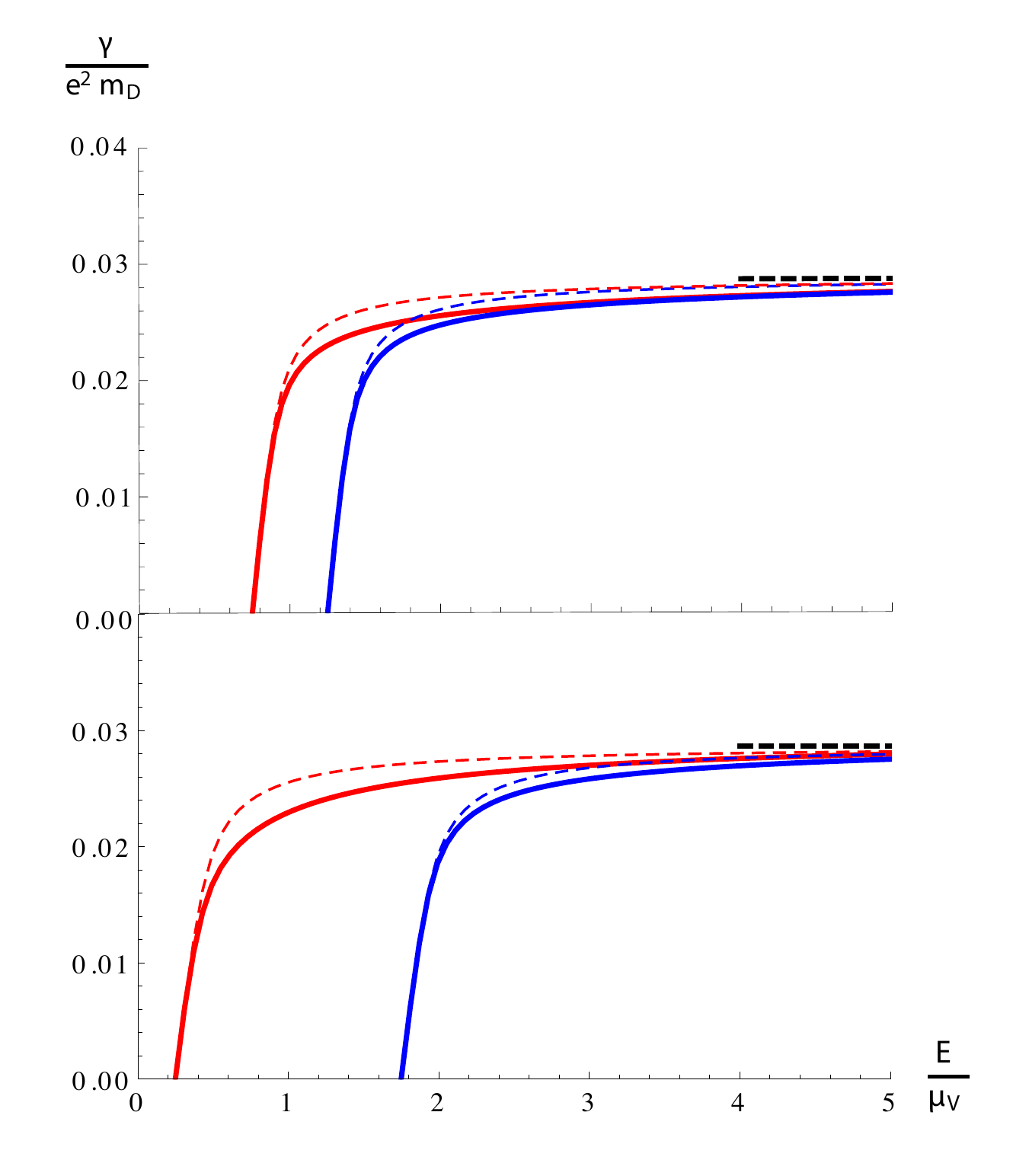}
\caption{Damping curves for left-  (red lines) and right-handed (blue lines) fermions for a fixed ratio $\mu_5/\mu_V = 1/4$ (top) and $3/4$ (bottom). The solid lines denote the result of the full numerical integration of Eq.~(\ref{generalexp})  up to order $1/E$, while
the dashed lines show for comparison only the leading-order results without $1/E$ corrections. The curves for both chiralities approach the limit $\gamma \sim 0.027 \,e^2 m_D$,
which is denoted as a black dashed line \label{fig:dampnum} }
\end{figure}
\end{center}

When extracting the leading contribution close to the Fermi surface, we find the same leading behavior as in chiral symmetric matter \cite{LeBellac:1996kr,Vanderheyden:1996bw,Gerhold:2005uu}, namely
\beq
\gamma_\chi(E) = \frac{e^2}{24\pi} (E-\mu_\chi) +{ \cal{O}}\Big((E-\mu_\chi)^{5/3}\Big) \, .
\eeq 
 As a by-product we can say that non-Fermi liquid behaviour is then expected for the two  chiral fermion populations.
In Fig.\ref{fig:dampnum} we show the result of the numerical integration of Eq.~(\ref{generalexp})  for left- and right-handed fermions  for two fixed values of the ratio $\mu_5/\mu_V$ (we choose $\mu_5 >0$ so that the density of right-handed fermions is larger than the left-handed ones).
The damping is zero at their respective Fermi surfaces ($E = \mu_{R,L}$)  then grows with the energy and  approaches the same limit for both chiralities. For a fixed value of the energy the value of the damping for left-handed fermions is always larger, indicating that for $\mu_5>0$ right-handed fermions have a longer lifetime compared to the left-handed ones.
The inclusion of $1/E$ effects provides a visible effect as we move away from the Fermi surface: more specifically, for a positive $\mu_5$ the damping rate for a left-handed fermion increases more slowly with energy than its right-handed counterpart. While being subleading, this effect becomes stronger as $\mu_5$ increases, as can be seen from the figure.

 Let us comment at this point on the regime of validity of our result.  The fact that the damping rate vanishes at the Fermi surface has been concluded from a perturbative computation. This result can be
 understood in terms of phase space restrictions and Pauli blocking, as  the imaginary part of the fermion self-energy can be written as a decay rate \cite{Weldon:1983jn}. 
It is then possible to see that
there is no phase space available for the scattering of a fermion at the Fermi surface with one in the Fermi sea producing two final states
above the Fermi sea. As the inverse of the damping rate gives the lifetime of the quasiparticle, one could apparently conclude the stability of the Fermi sea.
However, when the unstable gauge modes grow large, at time scales $t_{\rm ins} \sim 1/(\alpha^2 \mu_5)$ \cite{Akamatsu:2013pjd}, perturbation theory ceases to be valid.
Thus, we can only conclude that the Fermi sea is stable at shorter time scales.

\section{Conclusions} 
\label{conclusions}

In this work, we have shown how to  extend the HTL resummation scheme to a plasma in the presence of a chiral imbalance,
and discussed how the dispersion relations for the collective modes are altered by the presence of a chiral chemical potential. 
Aside from the appearance of imaginary poles which may give rise to plasma instabilities, we investigated in detail the splitting of the  collective modes associated 
with the different circular polarizations of the photons. 

We then
computed the damping rate of a fermion in a chiral plasma
 by taking into account effects of the chiral imbalance
both in the fermion and the resummed photon propagators.
 The value of the damping rate was found to depend on the chirality of the fermion, unless its energy
 is much larger than  the chemical potentials of the system, when a unique asymptotic value is reached.
 It is important to recall that 
our results should be valid  for time scales shorter than the onset of the instability, as for larger times a perturbative analysis as the
one we have carried out ceases to be valid.
 For larger time scales one
 should then derive an effective field theory to describe the system at those time scales in order to see how the chiral plasma instability affects the Fermi seas, similar
to the approaches carried out to see the evolution of Weibel instabilities for non- Abelian plasmas in Ref.\cite{Rebhan:2004ur}. At finite temperature, and at time scales where the hydrodynamical regime is reached one can actually study the dynamical evolution of the chiral chemical potential 
by studying Maxwell's equations combined with the chiral anomaly equation \cite{Boyarsky:2011uy,Manuel:2015zpa}.

It would  also be interesting to compute the fermion lifetime for non-vanishing values of the temperature. As for the chiral symmetric case, the Bose enhancement associated with the
photon propagator in the fermion self-energy yields an infrared divergence, and further resummations of Feynman diagrams would then be required \cite{Blaizot:1996hd}.
 Furthermore, such an additional infrared enhancement together with the additional structure generated by the chiral imbalance might introduce additional issues of non-integrability, and a careful 
investigation on whether the dynamical shielding still holds would be required.

 Finally, while we have seen that already for a QED plasma  the presence of a chiral imbalance induces significant new effects,
it would also be interesting to explore the case of non-Abelian (QCD or electroweak) plasmas, especially for their cosmological implications. 
 In the non-Abelian case, in order to preserve gauge invariance the resummation techniques not
only modify the gauge propagators but also the vertex functions whenever the external lines have soft momenta.
 The breaking of the $P$ and $CP$ symmetries suggests that the vertices would be
different for the different gauge transverse polarized  states,  leading to different imprints of the chiral fermion imbalance.

{\bf Acknowledgments:}
We thank J. Soto and J. Torres-Rincon for useful discussions.
We have been supported by the MINECO (Spain) under the projects  FPA2016-81114-P and  FPA2016-76005-C2-1-P,
as well as by the
project  2017-SGR-929  (Catalonia).
This work was also supported by the COST Action CA15213 THOR.

\end{document}